\documentclass[11pt]{article}

\usepackage{amsmath,amsfonts,amssymb}
\usepackage{graphicx}
\usepackage{hyperref}
\usepackage{authblk}
\usepackage{geometry}
\usepackage{tikz}
\usepackage{caption}
\usepackage{fancyhdr}
\usepackage{pifont}
\usepackage{amssymb}
\usepackage{pifont}  

\usepackage[numbers,sort&compress]{natbib}  

\usepackage[T1]{fontenc}   
\usepackage{upquote}       

\usepackage[most]{tcolorbox}
\tcbuselibrary{listings, breakable}

\tcbset{
  decoratorverbatim/.style={
    enhanced,
    breakable,
    colback=gray!5,      
    colframe=gray!40,    
    boxrule=0.5pt,
    arc=2pt,
    outer arc=2pt,
    left=4pt,
    right=4pt,
    top=4pt,
    bottom=4pt,
    fontupper=\ttfamily,
  }
}

\usepackage{xcolor}  
\usepackage{booktabs}
\usepackage{tabularx} 

\usepackage{tikz}
\usetikzlibrary{arrows.meta,positioning,fit,calc,backgrounds,shapes.misc}
\usepackage{amsmath, amssymb, bm}
\usepackage[export]{adjustbox}

\usepackage{tikz}
\usetikzlibrary{arrows.meta, positioning}
\usetikzlibrary{arrows.meta, positioning, shapes.multipart}
\usetikzlibrary{shapes.geometric, positioning}
\usepackage{amsmath,amssymb,amsfonts}
\usepackage{graphicx}
\usepackage{hyperref}
\usepackage{enumitem}
\usepackage{booktabs}
\usepackage{bm}

\usepackage{booktabs}
\usepackage{multirow}
\usepackage{pdflscape}     

\usepackage{tikz}
\usetikzlibrary{arrows.meta,positioning,matrix}
\usetikzlibrary{positioning}

\title{\textbf{Prompt Decorators: A Declarative and Composable Syntax for Reasoning, Formatting, and Control in LLMs}}
\author{
\textbf{Mostapha Kalami Heris} \thanks{Correspondence: \texttt{m.k.heris@shu.ac.uk}} \\
School of Engineering and Built Environment \\
College of Business, Technology and Engineering \\
Sheffield Hallam University
}

\date{October 21, 2025}

\begin{document}

\maketitle

\begin{abstract}
Large Language Models (LLMs) are central to reasoning, writing, and decision-support workflows, yet users lack consistent control over how they reason and express outputs. Conventional prompt engineering relies on verbose natural-language instructions, limiting reproducibility, modularity, and interpretability. This paper introduces \textit{Prompt Decorators}---a declarative, composable syntax that governs LLM behavior through compact control tokens such as \texttt{+++Reasoning}, \texttt{+++Tone(style=formal)}, and \texttt{+++Import(topic="Systems Thinking")}. Each decorator modifies a behavioral dimension, such as reasoning style, structure, or tone, without changing task content. The framework formalizes twenty core decorators organized into two functional families (\textit{Cognitive \& Generative} and \textit{Expressive \& Systemic}), each further decomposed into subcategories that govern reasoning, interaction, expression, and session-control. It defines a unified syntax, scoping model, and deterministic processing pipeline enabling predictable and auditable behavior composition. By decoupling task intent from execution behavior, Prompt Decorators create a reusable and interpretable interface for prompt design. Illustrative use cases demonstrate improved reasoning transparency, reduced prompt complexity, and standardized model behavior across domains. The paper concludes with implications for interoperability, behavioral consistency, and the development of declarative interfaces for scalable AI systems.
\end{abstract}


\section{Introduction}
\label{sec:introduction}

Large Language Models (LLMs) such as GPT, Claude, Gemini, and LLaMA have become foundational tools in modern computational workflows, supporting reasoning, writing, programming, and decision-making across domains \cite{schulhoff2025promptreport, bommasani2021opportunities}. Their increasing adoption has transformed how knowledge work is mediated, yet the challenge of controlling \textit{how} these models behave persists. Current prompting practices rely primarily on verbose and inconsistent natural-language instructions to specify reasoning depth, tone, and structure. While this approach is intuitive, it lacks standardization, reproducibility, and transparency—leading to unpredictable or inconsistent results across sessions and models \cite{vatsal2024survey}.

Natural language, though expressive, is an unreliable medium for procedural control. Small syntactic or lexical variations can cause large behavioral divergences in model reasoning and output style \cite{schulhoff2025promptreport}. As LLMs are deployed in high-stakes and multi-agent environments—ranging from analytical decision support to software generation—the absence of a formal mechanism for behavioral specification emerges as a critical limitation. Users seeking reliability, traceability, and interpretability must rely on trial-and-error phrasing rather than a transparent or declarative interface \cite{beurerkellner2023lmql}.

To address this gap, this paper introduces \textbf{Prompt Decorators}, a declarative and composable syntax for specifying how models reason, format, and interact—without altering the semantic content of prompts. Each decorator represents a compact behavioral directive expressed in a structured but human-readable form. For example:

\begin{tcolorbox}[decoratorverbatim]
\begin{verbatim}
+++Reasoning
+++Debate

Explain the implications of using facial recognition in public spaces.
\end{verbatim}
\end{tcolorbox}

This prompt instructs the model to (1) articulate its reasoning explicitly before providing conclusions, and (2) format the output in Markdown. Behavioral intent is thereby encoded concisely through declarative symbols rather than embedded natural-language instructions.

Prompt Decorators introduce three key contributions to the practice of LLM interaction:

\begin{enumerate}
    \item \textbf{Declarativity:} Enables users to define desired reasoning and output behaviors explicitly, decoupling behavioral intent from linguistic phrasing.
    \item \textbf{Composability:} Allows multiple decorators to be stacked modularly, forming reusable configurations for reasoning style, tone, or structural control.
    \item \textbf{Transparency:} Exposes behavioral logic in an explicit and inspectable format, enhancing reproducibility and interpretability across sessions.
\end{enumerate}

The framework formalizes twenty core decorators grouped into two functional families (\textit{Cognitive \& Generative} and \textit{Expressive \& Systemic}). These families are decomposed into subcategories that govern reasoning, interaction, expression, and session-control. Collectively, they define a standardized vocabulary for managing reasoning, tone, structure, and conversational persistence within LLM interfaces; see Figure~\ref{fig:prompt-taxonomy} and Table~\ref{tab:decorator-taxonomy}.

By separating task semantics from behavioral directives, Prompt Decorators move prompt design from an informal linguistic craft toward a structured, auditable interface. This paradigm aligns with ongoing research into declarative and structured prompting frameworks such as LMQL \cite{beurerkellner2023lmql}, BAML \cite{boundaryml2024baml}, and DSPy \cite{khattab2024dspy, khattab2022demonstrate}. In doing so, it establishes a foundation for reproducible prompt engineering, transparent reasoning control, and scalable behavioral governance across applications.

The remainder of this paper is organized as follows. 
Section~\ref{sec:related-work} reviews prior research on prompt engineering, structured prompting frameworks, and declarative control in LLM interfaces. 
Section~\ref{sec:decorator-framework} introduces the Prompt Decorators Framework, detailing its functional families, subcategories, and representative decorators. 
Section~\ref{sec:methodology} describes the research design, formalization process, and evaluation and validation methods. 
Section~\ref{sec:use-cases} presents six applied use cases and distills cross-case insights and design principles. 
Section~\ref{sec:risks} discusses risks and limitations. 
Section~\ref{sec:future-work} outlines future work and possible extensions. 
Finally, Section~\ref{sec:conclusion} concludes the paper.


\section{Related Works and Background}
\label{sec:related-work}

\subsection{Prompt Engineering}

Prompt engineering has become a cornerstone of adapting Large Language Models (LLMs) to diverse downstream applications, including reasoning, analysis, and creative synthesis. Unlike conventional software systems, which rely on explicit parameters or APIs, LLMs expose their functionality primarily through natural-language interfaces. Consequently, prompt formulation acts as a high-level programming layer that governs both task performance and model interpretability \cite{schulhoff2025promptreport, vatsal2024survey, bommasani2021opportunities}.

Recent meta-analyses have brought increasing conceptual structure to this emerging discipline. Schulhoff et al. \cite{schulhoff2025promptreport} conducted the most comprehensive taxonomy to date, classifying over fifty prompting paradigms into five principal categories—\textit{In-Context Learning}, \textit{Thought Generation}, \textit{Decomposition}, \textit{Self-Criticism}, and \textit{Prompt Ensembling}. These categories extend the notion of prompting beyond instruction following, introducing iterative reasoning, error analysis, and compositional workflows. Similarly, Vatsal and Dubey \cite{vatsal2024survey} surveyed 39 prompting methods across 29 NLP tasks, emphasizing the increasing role of structured reasoning strategies such as Chain-of-Thought (CoT), ReAct, and Tree-of-Thoughts (ToT). Despite this proliferation, the field remains largely procedural—dependent on linguistic phrasing, implicit heuristics, and user-driven trial-and-error rather than formal behavioral specification.

Prompting effectiveness typically depends on three functional roles: (1) guiding reasoning processes, (2) structuring the generated outputs, and (3) aligning expression with contextual intent. However, studies consistently report sensitivity to minor lexical or syntactic variations, which can yield divergent reasoning trajectories or answer formats \cite{schulhoff2025promptreport}. This instability raises fundamental issues of reproducibility and behavioral auditability. The same surveys highlight three recurrent failure modes: (a) \textit{prompt sensitivity}—variance in output under semantically equivalent phrasings; (b) \textit{context interference}—order effects in in-context examples; and (c) \textit{semantic drift}—behavioral degradation under iterative reformulation. These pathologies underscore the limits of natural language as a procedural control surface.

To address these issues, prior research converges on three dominant paradigms of practice:

\begin{enumerate}
    \item \textbf{Instructional Prompting:} Encodes task goals and stylistic constraints directly in natural language (e.g., “Write a 200-word summary in a formal tone”). This method is intuitive but verbose, inconsistent, and prone to ambiguity across paraphrases \cite{vatsal2024survey}.
    \item \textbf{Reasoning Prompting:} Structures intermediate cognitive steps using explicit reasoning chains such as CoT \cite{wei2022chainofthought}, Self-Consistency \cite{wang2022selfconsistency}, and ReAct \cite{yao2022react}. These approaches improve interpretability but remain descriptive of \textit{how} reasoning unfolds rather than declarative of \textit{what reasoning behavior} should be enforced.
    \item \textbf{Meta-Prompting:} Treats prompts as modular and composable artifacts. Systems such as AutoPrompt, PromptLayer, and PromptChainer introduce parameterization, chaining, and versioning at the application level \cite{schulhoff2025promptreport}, yet their functionality operates outside the conversational layer, limiting accessibility for non-developers.
\end{enumerate}

Across these paradigms, behavioral semantics remain embedded within natural language. This entanglement constrains modularity, reusability, and formal reasoning about LLM behavior. These limitations motivate a shift toward declarative abstractions that separate behavioral intent from task semantics. The proposed \textbf{Prompt Decorators} address this gap by defining compact, interpretable behavioral tokens that operate natively within natural-language interfaces while preserving formal control semantics.

\subsection{Structured Prompting Frameworks}

In response to the brittleness of unstructured natural-language prompts, recent work has introduced frameworks that formalize LLM interaction as programmable or declarative processes. These systems extend prompting into structured or hybrid query languages that support constraints, composability, and validation.

\textbf{LMQL (Language Model Query Language)} \cite{beurerkellner2023lmql} reframes prompting as a query execution process that integrates variables, constraints, and control flow into a unified syntax. By compiling natural-language segments into executable queries, LMQL enables conditional generation and constraint enforcement while reducing inference cost by 26–85\%. Its hybrid model demonstrates that structured control can increase both efficiency and determinism.

\textbf{BAML (Behavioral API Modeling Language)} \cite{boundaryml2024baml} takes a software-engineering perspective, formalizing model interactions as typed behavioral APIs. It defines explicit schemas for input-output mappings, enabling static validation and reproducibility across environments. BAML thus transforms prompts into declarative behavioral contracts that integrate with existing application backends.

\textbf{DSPy (Declarative Structured Programming for LLMs)} \cite{khattab2024dspy, khattab2022demonstrate} generalizes this direction by introducing differentiable prompt modules optimized through feedback and learning signals. DSPy abstracts reasoning templates into composable functional components, which can be trained and tuned dynamically to achieve desired behavior.

While these frameworks advance the control and reproducibility of LLM systems, they remain primarily developer-oriented. Their abstractions—expressed in code or configuration—optimize workflow and pipeline management rather than end-user behavioral semantics. Consequently, non-programmatic users lack a direct linguistic mechanism for controlling reasoning depth, tone, or output structure within conversational contexts.

\textbf{Prompt Decorators} complement these frameworks by introducing a lightweight, human-readable syntax that is also formally interpretable. They bridge the gap between natural language and structured control, allowing declarative behavioral directives to coexist with conversational expressiveness. This enables decorators to serve as an interface layer compatible with LMQL, BAML, and DSPy, while remaining accessible to non-technical users.

\subsection{Declarative Control in LLM Interfaces}

Declarative systems shift emphasis from procedural description (\textit{how} to act) to behavioral specification (\textit{what} to achieve). Within the LLM domain, declarativity enables the articulation of behavioral requirements—such as reasoning transparency, tone regulation, or format consistency—without prescribing implementation details.

\textbf{Prompt Decorators} extend this paradigm by defining a human-readable yet machine-interpretable syntax that encodes behavioral logic explicitly. Each decorator can be conceptualized as a tuple $\langle s, p, c \rangle$ representing a \textit{scope} $s$, a \textit{parameterization} $p$, and a \textit{control intent} $c$, which together form a declarative constraint on the model’s generative process. For instance, \texttt{+++Reasoning} specifies the inclusion of intermediate thought steps, while \texttt{+++Tone(style=formal)} defines expressive constraints. These declarative tokens can be composed, nested, or scoped, creating a higher-order behavioral grammar over LLM interaction.

Unlike procedural prompts that embed behavioral cues within natural-language content, or programmatic frameworks that externalize prompts into code, Prompt Decorators unify both. They provide:

\begin{itemize}
    \item A linguistically natural yet formally structured control syntax.
    \item A composable mechanism for managing reasoning depth, tone, and output format.
    \item Transparent and auditable behavioral semantics compatible with structured frameworks.
\end{itemize}

This approach advances prompt engineering from ad hoc linguistic experimentation toward a reproducible and interpretable interaction paradigm. By making behavioral control explicit, Prompt Decorators also contribute to governance and safety: they enable clearer documentation of prompt logic, facilitate behavior audits, and reduce vulnerabilities to prompt injection \cite{schulhoff2025promptreport}. Conceptually, they represent the next stage in the evolution of LLM interaction design—progressing from implicit natural-language control through structured programming to fully declarative behavioral specification.


\section{The Prompt Decorators Framework}
\label{sec:decorator-framework}

This section introduces the \emph{Prompt Decorators} framework—a declarative system for controlling how a Large Language Model (LLM) reasons, structures, and expresses its responses without changing the task itself. 
Prompt Decorators function as modular directives that specify \textit{how} to think or respond rather than \textit{what} to discuss. 
We outline the underlying concept, syntax and scope, processing pipeline, and representative examples.

\subsection{Concept and Motivation}

Prompt Decorators operate as behavioral modifiers that wrap user prompts to make model behavior explicit, reproducible, and inspectable. 
Inspired by Python decorators, which wrap functions to alter execution, Prompt Decorators wrap instructions to influence reasoning style, interaction pattern, and presentation. 
This separation between task intent and execution behavior enables more deliberate, auditable control over LLM outputs.

For instance, a Python decorator can wrap a function to add pre- and post-processing logic without changing the function's core definition:

\begin{tcolorbox}[decoratorverbatim]
\begin{verbatim}
def log_execution(func):
    def wrapper(*args, **kwargs):
        print("Starting execution...")
        result = func(*args, **kwargs)
        print("Execution completed.")
        return result
    return wrapper

@log_execution
def compute():
    print("Running computation.")

compute()
\end{verbatim}
\end{tcolorbox}

In this example, the decorator \texttt{@log\_execution} modifies the behavior of the \texttt{compute()} function by inserting logging steps before and after its execution, without altering the function's internal logic. 
This illustrates the principle of declarative wrapping—where additional behavior is composed around a core intent—forming the conceptual basis for Prompt Decorators.

Similarly, Prompt Decorators wrap user instructions to modify the reasoning process or expressive behavior while leaving the core query intact:

\begin{tcolorbox}[decoratorverbatim]
\begin{verbatim}
+++Rewrite
+++Reasoning

Explain the implications of using generative AI in education.
\end{verbatim}
\end{tcolorbox}

In this example, \texttt{+++Rewrite} instructs the model to reinterpret and refine the given prompt for clarity before generating a response, 
while \texttt{+++Reasoning} requests that the model produce an explicit logical explanation prior to the final answer. 
When composed, these decorators direct the model to first reformulate the task, then reason through it transparently, 
demonstrating how declarative composition can shape the cognitive sequence of model behavior without altering the underlying query.

\subsection{Syntax and Scoping}

\paragraph{Syntax.}
Each decorator follows a compact, machine-interpretable structure:

\begin{tcolorbox}[decoratorverbatim]
\begin{verbatim}
+++Name(optional_parameters)
\end{verbatim}
\end{tcolorbox}

A decorator consists of a canonical name, optional key–value parameters, and sequential composition. 
Multiple decorators may be stacked, processed from top to bottom in a predictable order. 
This simple form supports both human readability and potential programmatic parsing.

\paragraph{Scoping.}
Decorators can apply locally or persist across turns:
\begin{itemize}
    \item \textbf{Message Scope} (\texttt{+++MessageScope}) — applies only to the current message.
    \item \textbf{Chat Scope} (\texttt{+++ChatScope}) — persists until explicitly cleared.
\end{itemize}

Utility decorators manage this state: \texttt{+++Clear} removes active chat-level decorators, 
\texttt{+++ActiveDecs} lists currently active ones, 
and \texttt{+++AvailableDecs} displays all supported decorators. 
Explicit scoping ensures that behavioral context remains visible and controllable throughout an interaction.

\subsection{Processing Model}

Prompt Decorators follow a layered processing pipeline that translates symbolic controls into structured behavior:

\begin{enumerate}
    \item \textbf{Parsing} — extract decorator names, parameters, and order.
    \item \textbf{Scope Resolution} — merge chat and message scopes, applying any clearing rules.
    \item \textbf{Planning and Interaction} — optionally plan, clarify, or import conceptual context.
    \item \textbf{Reasoning and Generation} — perform structured analysis, debate, critique, or refinement.
    \item \textbf{Formatting and Expression} — apply tone, style, and output constraints.
    \item \textbf{Introspection and Export} — reveal current state or export metadata for auditing.
\end{enumerate}

This pipeline mirrors middleware architectures: each stage incrementally refines behavior, ensuring consistency between declared intent and model execution.

\subsection{Selected Decorators}

Prompt Decorators constitute a structured vocabulary of declarative controls that shape how Large Language Models (LLMs) reason, plan, critique, and express responses. Each decorator operates independently yet can be composed with others, forming reproducible behavioral configurations. Table~\ref{tab:decorator-taxonomy} summarizes all defined Prompt Decorators, organized by their functional categories and subcategories. It provides a concise overview of how cognitive, generative, expressive, and systemic decorators collectively structure reasoning and behavioral control. As visualized in Figure~\ref{fig:prompt-taxonomy}, the framework groups decorators into two main functional families—Cognitive \& Generative and Expressive \& Systemic—each decomposed into subcategories that govern reasoning, interaction, expression, and meta-control behaviors.

Below, we describe representative decorators that illustrate different functional families within the framework. Each entry details its intent, behavioral pattern, and typical use. Full technical definitions, parameters, and examples are available in the official repository.

\noindent
GitHub Repository: \url{https://github.com/smkalami/prompt-decorators}

\begin{figure}[t]
\centering
\includegraphics[width=\linewidth]{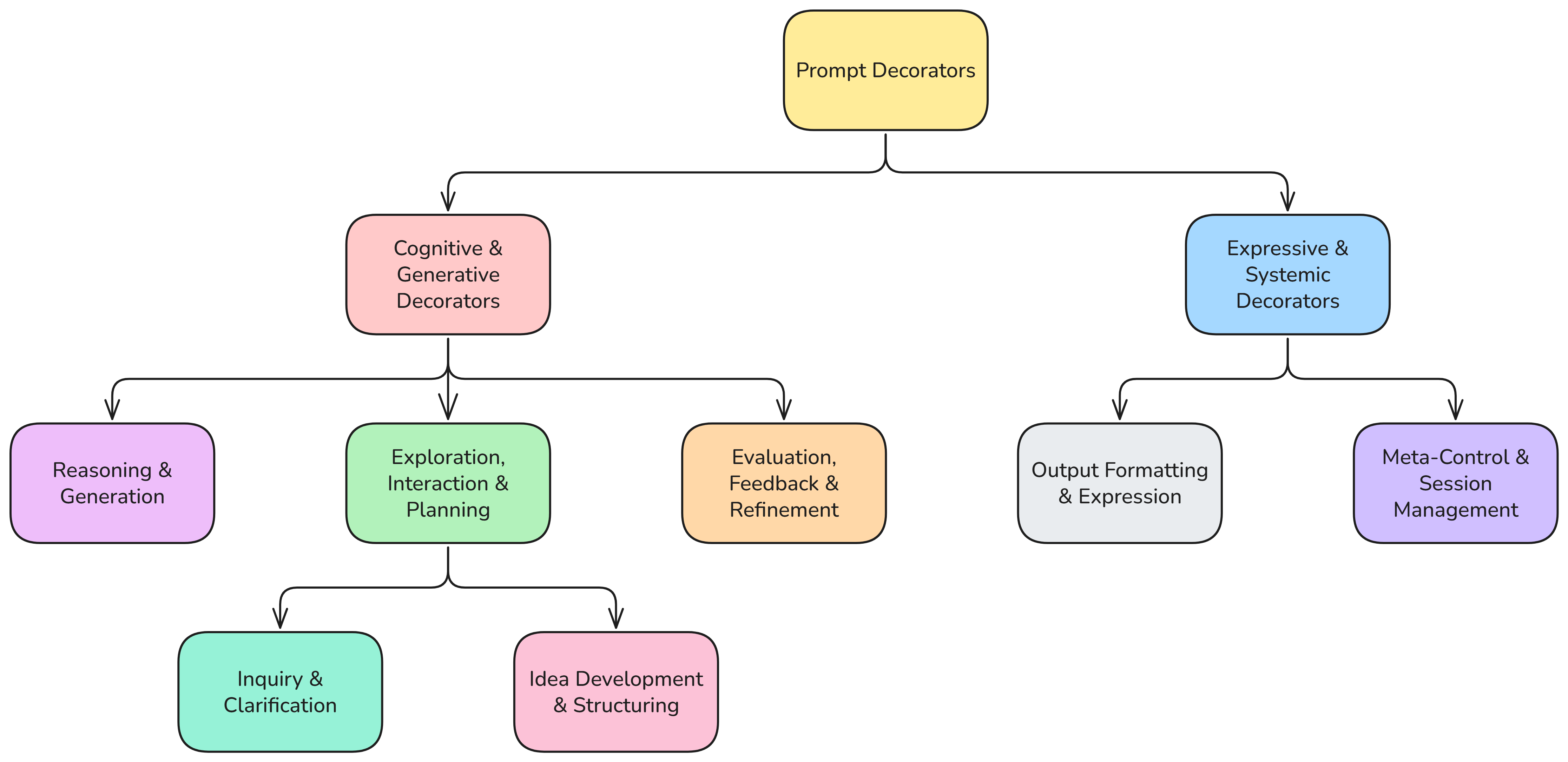}
\caption{Hierarchical taxonomy of \textit{Prompt Decorators}, organized into two primary functional families, Cognitive \& Generative and Expressive \& Systemic, each decomposed into subcategories that govern reasoning, interaction, expression, and meta-control behaviors. The diagram illustrates how decorators collectively span the cognitive, expressive, and systemic layers of model behavior.}
\label{fig:prompt-taxonomy}
\end{figure}


\begin{landscape}

\begin{table}[htbp]
\centering
\small
\caption{Taxonomy of Prompt Decorators with Functional Categories and Subcategories}
\label{tab:decorator-taxonomy}
\begin{tabular}{p{4.5cm} p{4.5cm} p{3cm} p{9cm}}
\toprule
\textbf{Category} & \textbf{Subcategory} & \textbf{Decorator} & \textbf{Function / Description} \\
\midrule
\multirow{8}{*}{\textbf{Cognitive \& Generative}} 
 & Reasoning \& Generation & \texttt{+++Reasoning} & Provide reasoning before final answer to improve transparency and traceability. \\
 &  & \texttt{+++StepByStep} & Execute the task in labeled steps with a final synthesis. \\
 &  & \texttt{+++Debate} & Present multiple positions before synthesizing a conclusion. \\
\cmidrule(lr){2-4}
 & Inquiry \& Clarification & \texttt{+++Interactive} & Ask clarification questions when prompt is underspecified. \\
 &  & \texttt{+++Socratic} & Apply Socratic questioning to surface assumptions and deepen understanding. \\
\cmidrule(lr){2-4}
 & Planning \& Ideation & \texttt{+++Planning} & Outline plan and objectives before task execution. \\
 &  & \texttt{+++Brainstorm} & Generate multiple labeled ideas without judgment. \\
 &  & \texttt{+++Rewrite} & Reframe the user prompt into a clearer or more actionable version. \\
 &  & \texttt{+++Import} & Import a conceptual lens or discipline into reasoning. \\
\cmidrule(lr){2-4}
 & Evaluation \& Feedback & \texttt{+++Critique} & Provide structured feedback with strengths, weaknesses, and improvements. \\
 &  & \texttt{+++Refine} & Iteratively improve the output through labeled passes. \\
 &  & \texttt{+++Candor} & Control directness and bluntness of feedback. \\
\midrule
\multirow{7}{*}{\textbf{Expressive \& Systemic}} 
 & Output Formatting & \texttt{+++OutputFormat} & Enforce syntactically valid output structure (JSON, YAML, Markdown, etc.). \\
 &  & \texttt{+++Tone} & Configure tone or stylistic register (formal, technical, friendly, etc.). \\
\cmidrule(lr){2-4}
 & Session \& Meta Control & \texttt{+++ChatScope} & Activate persistent behavior across conversation turns. \\
 &  & \texttt{+++MessageScope} & Restrict decorator effects to the current message only. \\
 &  & \texttt{+++Clear} & Remove all or selected decorators from chat scope. \\
 &  & \texttt{+++ActiveDecs} & List all active decorators in the current chat session. \\
 &  & \texttt{+++AvailableDecs} & Display catalog of supported decorators with activation status. \\
 &  & \texttt{+++Export} & Export conversation content and metadata for auditing or recordkeeping. \\
\bottomrule
\end{tabular}
\end{table}

\end{landscape}


\paragraph{\colorbox{gray!20}{\texttt{+++Reasoning}}}
This decorator exposes a structured reasoning section before the final answer, enhancing transparency and interpretability. When active, the model first presents its logic or assumptions in a labeled \textit{Reasoning} block, followed by a clearly marked \textit{Final Answer}. It is particularly useful for analytical writing, academic discussion, or decision-making contexts where justifications are essential.

\paragraph{\colorbox{gray!20}{\texttt{+++StepByStep}}}
This decorator decomposes the problem into labeled steps, enforcing ordered and logical progression. Each step is numbered (\textit{Step 1, Step 2, ...}) and ends with a concise \textit{Final Step} or summary. It is suitable for procedural reasoning, algorithmic explanations, and problem-solving scenarios that benefit from visible intermediate stages.

\paragraph{\colorbox{gray!20}{\texttt{+++Debate}}}
This decorator simulates balanced argumentation among multiple viewpoints before drawing a conclusion. It presents opposing or complementary positions labeled as \textit{Position A}, \textit{Position B}, etc., followed by a reasoned synthesis when enabled. It is valuable in ethics, policy, or strategy analyses that require structured dialectical reasoning.

\paragraph{\colorbox{gray!20}{\texttt{+++Interactive}}}
When the user prompt lacks essential context, this decorator triggers clarifying questions before proceeding. The model identifies ambiguity, asks targeted questions, and continues the task once answers are received. It mitigates hallucination risks and enhances task fidelity. It is especially useful in iterative workflows or technical documentation.

\paragraph{\colorbox{gray!20}{\texttt{+++Socratic}}}
Inspired by the Socratic method, this decorator guides the model to restate the question, surface assumptions, and probe through layered inquiry before providing synthesis. It encourages conceptual depth, reflection, and learning-oriented dialogue.

\paragraph{\colorbox{gray!20}{\texttt{+++Planning}}}
This decorator introduces a planning phase before execution. The model begins by outlining a brief plan that lists objectives, steps, and constraints, labeled as \textit{Plan}, followed by the main response under \textit{Execution}. It improves the coherence of complex or technical outputs and is well suited for research writing, project proposals, and system design tasks.

\paragraph{\colorbox{gray!20}{\texttt{+++Brainstorm}}}
Designed for creative exploration, this decorator directs the model to generate multiple labeled ideas or options without premature evaluation. Options are numbered and presented for variety and divergence, which can later be refined or critiqued by additional decorators such as \texttt{+++Critique} or \texttt{+++Refine}.

\paragraph{\colorbox{gray!20}{\texttt{+++Rewrite}}}
This decorator instructs the model to reformulate the user’s original prompt into a clearer, more actionable version before generating a response. The output is typically structured in two segments: \textit{Rewritten Prompt}, which presents the clarified formulation, and \textit{Response Based on Rewritten Prompt}, which provides the corresponding answer. By reducing ambiguity, enforcing structure, and standardizing task interpretation, \texttt{+++Rewrite} enhances clarity and response fidelity—functioning as an embedded prompt engineer that improves the interpretive precision of large language models.

\paragraph{\colorbox{gray!20}{\texttt{+++Import}}}
This decorator introduces a conceptual or disciplinary lens into the model’s reasoning process. 
The model explicitly identifies the imported concept and applies it consistently throughout the response. 
For example, \texttt{+++Import(topic="Systems Thinking")} directs the analysis to adopt systems-oriented logic and perspective. Conceptually, this decorator functions like importing a library in a programming language—activating a specialized set of interpretive routines without altering the underlying architecture. 
By conditioning the model’s representational focus, \texttt{+++Import} enhances interdisciplinary consistency, contextual depth, and reproducibility across sessions.

\paragraph{\colorbox{gray!20}{\texttt{+++Critique}}}
This decorator provides balanced, constructive feedback by identifying strengths, weaknesses, and actionable improvements. The model follows a structured pattern: \textit{Identify Subject} → \textit{Highlight Strengths} → \textit{Critique Weaknesses} → \textit{Suggest Improvements} → \textit{Conclude}. It is ideal for reviewing drafts, plans, or analytical arguments, especially when tone control and professionalism are required.

\paragraph{\colorbox{gray!20}{\texttt{+++Refine}}}
This decorator enables iterative enhancement by generating multiple labeled passes (\textit{Iteration 1}, \textit{Iteration 2}, ...), each improving clarity, coherence, or style until a \textit{Final Answer} is reached. The parameter \texttt{iterations=N} specifies the number of refinement cycles. It is highly effective for self-editing or progressive output optimization.

\paragraph{\colorbox{gray!20}{\texttt{+++Candor}}}
Controls the directness and bluntness of feedback while preserving professional tone. The parameter \texttt{level} can be set to \texttt{low} (diplomatic), \texttt{medium} (balanced), or \texttt{high} (blunt). It is often combined with \texttt{+++Critique} or \texttt{+++Tone} to match organizational culture or audience expectations.

\paragraph{\colorbox{gray!20}{\texttt{+++OutputFormat}}}
Enforces the final output to strictly follow a specified structure such as JSON, YAML, Markdown, or XML. It ensures syntactic validity and downstream compatibility with tools or pipelines. The decorator is typically used for structured outputs, documentation, or API integration, where consistent format adherence is essential.

\paragraph{\colorbox{gray!20}{\texttt{+++Tone}}}
Defines the stylistic register of the model’s expression—formal, casual, technical, friendly, or humorous—without altering factual content. It adjusts vocabulary, phrasing, and rhythm to match context and audience, and can be combined with \texttt{+++Candor} or \texttt{+++OutputFormat} for nuanced delivery control.

\paragraph{\colorbox{gray!20}{\texttt{+++ChatScope}}}
This decorator activates persistent behavior across the entire conversation rather than a single message. When \texttt{+++ChatScope} is declared, all decorators listed within the same message become active at the chat level and remain in effect for subsequent turns until explicitly cleared. It is useful when continuity of reasoning style, tone, or formatting is required throughout a multi-turn dialogue. Persistent scoping enhances behavioral consistency but also demands careful management to avoid unintended carryover effects.

\paragraph{\colorbox{gray!20}{\texttt{+++MessageScope}}}
This decorator restricts behavioral influence to the current message only, temporarily overriding any active \texttt{ChatScope} settings. It is ideal for one-time configurations or exceptions that should not affect the global context. Once the message is completed, previously active chat-level decorators resume automatically. This local isolation mechanism supports flexibility without requiring manual resets.

\paragraph{\colorbox{gray!20}{\texttt{+++Clear}}}
This decorator deactivates persistent decorators that were previously set in \texttt{ChatScope}. When used without parameters, it removes all active decorators; when parameters are specified (for example, \texttt{+++Clear(+++Reasoning, +++Tone)}), it clears only those named. It is essential for resetting conversational state and maintaining predictable behavior across sessions. Clear functions as a governance utility to ensure explicit state transitions.

\paragraph{\colorbox{gray!20}{\texttt{+++ActiveDecs}}}
This diagnostic decorator lists all currently active decorators within the chat scope. When invoked, it outputs a concise, structured list that enumerates which behavioral controls remain in effect. If no decorators are active, it responds with the message “No active decorators.” This visibility improves transparency, enabling users to verify the current configuration before issuing further prompts.

\paragraph{\colorbox{gray!20}{\texttt{+++AvailableDecs}}}
This informational decorator produces a complete catalog of all supported decorators, each accompanied by a brief description and activation status. The output is typically rendered as a table with three columns: \textit{Name}, \textit{Description}, and \textit{Status}. Active decorators are marked as “Active,” while inactive ones are labeled “Inactive.” This function aids discoverability and serves as a built-in reference for system capabilities.

\paragraph{\colorbox{gray!20}{\texttt{+++Export}} / \colorbox{gray!20}{\texttt{+++Dump}}}
This decorator exports or summarizes the current chat session for review, documentation, or audit purposes. 
When invoked, it outputs the conversation content—along with active decorator metadata—in a user-specified format such as plain text, Markdown, or JSON. 
It is particularly useful for transparency, reproducibility, and governance applications where conversational provenance must be retained or shared. 
An alias, \texttt{+++Dump}, provides identical functionality and serves as a shorthand for rapid or ad hoc exports. 
Both \texttt{+++Export} and its alias are formally defined in the official GitHub repository of the Prompt Decorator Framework.

\medskip
The system-oriented decorators (\texttt{+++ChatScope}, \texttt{+++MessageScope}, \texttt{+++Clear}, \texttt{+++ActiveDecs}, \texttt{+++AvailableDecs}, and \texttt{+++Export}) provide state control, traceability, and accountability. They ensure that Prompt Decorator configurations remain explicit, inspectable, and reversible across extended interactions.

\subsection{Illustrative Composition}

\begin{tcolorbox}[decoratorverbatim]
\begin{verbatim}
+++ChatScope
+++Reasoning
+++Tone(style=formal)
+++OutputFormat(format=markdown)

Assess the ethical implications of AI-driven recruitment systems.
\end{verbatim}
\end{tcolorbox}

Here, chat scope persists behavior across turns, reasoning precedes the conclusion, tone is formal, and formatting is Markdown. The same configuration can be reused and audited, which improves consistency and reduces prompt drift.

\subsection{Design Principles and Implementation Notes}

The Prompt Decorators framework adheres to six guiding principles: declarativity, composability, transparency, predictability, reusability, and accessibility. Together, these principles shift prompt design from trial-and-error phrasing toward a structured and auditable interface with observable second-order effects.

\paragraph{Declarativity}
Behavioral intent is expressed symbolically rather than procedurally. Each decorator formalizes a distinct behavioral dimension—reasoning, tone, or structure—allowing the model’s generative process to be inspected and controlled without embedding verbose natural-language instructions.

\paragraph{Composability}
Decorators operate as modular units that can be layered predictably. Stacking directives such as \texttt{+++Reasoning} and \texttt{+++Tone(style=formal)} creates composite behavioral configurations while maintaining semantic separation between task content and control syntax.

\paragraph{Transparency and Predictability}
All active decorators remain explicit within the prompt text, making behavioral configuration visible and auditable. The processing pipeline enforces a deterministic top-to-bottom execution order that ensures predictable outcomes across compositions.

\paragraph{Reusability}
Decorator chains can be reused as configuration templates for recurring reasoning or stylistic contexts. This supports standardization across sessions, models, and collaborative workflows, reducing prompt drift and enabling behavioral reproducibility.

\paragraph{Accessibility}
While formally structured, the syntax remains human-readable. Users without programming expertise can declaratively manage model behavior using familiar textual patterns, bridging natural-language prompting and structured control.

\paragraph{Implementation Notes}
The current prototype implements the six-stage pipeline (parsing, scope resolution, planning, reasoning, formatting, introspection) described earlier in Section~3.3. Each decorator is processed as an independent function within a middleware layer that maps symbolic directives to corresponding behavioral routines. Scoping mechanisms (\texttt{+++ChatScope} and \texttt{+++MessageScope}) are realized through contextual state tracking, ensuring that persistent or local configurations remain interpretable and reversible. This implementation demonstrates that the declarative syntax can be operationalized within both conversational and programmatic environments.


\section{Methodology}
\label{sec:methodology}

This section outlines the methodological framing used to conceptualize, formalize, and validate the Prompt Decorators framework. While the system is primarily declarative rather than algorithmic, its development follows an explicit methodology integrating design-based reasoning, comparative analysis, and representational formalization—that is, the systematic encoding of conceptual distinctions into symbolic syntax.

\subsection{Research Design}
The approach follows a design science methodology aimed at developing an artifact that bridges human–AI interaction theory and practical control mechanisms. The design process iteratively balanced conceptual generality with practical interpretability through three cycles: (1) abstraction of recurring behavioral control patterns in LLM prompting, (2) synthesis of these patterns into a formal declarative syntax, and (3) validation through illustrative use cases and consistency analysis. Each cycle refined both semantic clarity and compositional expressiveness.

\subsection{Formalization Process}
To ensure methodological rigor, each decorator was defined as a behavioral primitive with explicit semantics, parameters, and composability rules. The framework employs a symbolic notation (\texttt{+++Name(parameter=value)}) to capture control intent independently from task semantics. A six-stage processing pipeline—parsing, scope resolution, planning, reasoning, formatting, and introspection—operationalizes these declarative instructions into structured model behaviors. This pipeline functions as an interpretive layer between user intent and LLM output.

\subsection{Evaluation and Validation}
Rather than empirical benchmarking, validation focuses on structured scenario analysis. Representative prompts were executed across reasoning, expressive, and systemic dimensions to assess consistency, modularity, and interpretability. Outputs were evaluated for (1) fidelity of behavioral intent, (2) stability under composition, and (3) interpretive transparency. This qualitative validation demonstrates that the declarative syntax can reproduce predictable behavioral configurations without procedural ambiguity.

\subsection{Methodological Contribution}
The methodological contribution of Prompt Decorators lies in formalizing prompt design as a reproducible, declarative process rather than an informal linguistic practice. By grounding behavioral control in structured syntax and interpretable processing, the framework transforms prompt engineering into a form of methodological inquiry—where reasoning transparency, compositional logic, and reproducibility are treated as measurable design properties. This establishes a reproducible foundation for behavioral control in human–LLM interaction.


\section{Applied Use Cases of Prompt Decorators}
\label{sec:use-cases}

Prompt Decorators can be composed to address recurring challenges in professional and research workflows that demand high reasoning quality, structural rigor, and interpretability. 
By declaratively specifying \textit{how} a model should deliberate, critique, or collaborate, users can create reproducible and auditable reasoning processes.
The following examples illustrate realistic and ethically grounded applications across diverse domains, demonstrating how decorators enhance cognitive transparency and task reliability.

\subsection{Multi-Perspective Feature Evaluation}

\textbf{Decorators:} \texttt{+++Debate}, \texttt{+++Reasoning}, and \texttt{+++Refine(iterations=3)}

\paragraph{Context and Pain Point}
Product teams frequently need to evaluate a new feature’s strategic value by weighing its benefits, risks, and implementation complexity.
Traditional analyses can be biased toward a single stakeholder viewpoint, limiting strategic objectivity.

\paragraph{Solution}
\texttt{+++Debate} generates structured contrasting viewpoints (for example, business opportunity versus technical feasibility). 
\texttt{+++Reasoning} enforces logical structure within each perspective, and \texttt{+++Refine(iterations=3)} progressively synthesizes these viewpoints into a coherent, justified recommendation.

\paragraph{Outcome}
The model produces a traceable decision rationale that captures trade-offs across multiple disciplines, enhancing accountability and strategic alignment.

\paragraph{Example}

\begin{tcolorbox}[decoratorverbatim]
\begin{verbatim}
+++Debate
+++Reasoning
+++Refine(iterations=3)

Evaluate whether to launch an AI-based summarization feature in beta.
\end{verbatim}
\end{tcolorbox}

\textbf{Output (abridged):} 
Outputs are shortened for brevity while retaining the logical structure and decorator interactions.

\begin{tcolorbox}[decoratorverbatim]
\begin{verbatim}
[Debate]
Perspective 1: Early launch accelerates user feedback and product learning.
Perspective 2: Premature release may harm brand trust if accuracy is low.

[Reasoning]
Balancing learning speed with quality assurance is key.

[Refinement 1]
Combine limited rollout with opt-in feedback collection.

[Refinement 2]
Focus beta on enterprise users and track satisfaction metrics.

[Refined Recommendation]
Proceed with a monitored beta phase limited to selected clients.
\end{verbatim}
\end{tcolorbox}

\subsection{Hypothesis Structuring Assistant}

\textbf{Decorators:} \texttt{+++Socratic}, \texttt{+++Planning}, and \texttt{+++StepByStep}

\paragraph{Context and Pain Point}
Researchers often begin with loosely defined ideas that require clarification into specific, testable hypotheses.
Without structured reasoning, this process can produce vague or logically inconsistent outcomes.

\paragraph{Solution}
\texttt{+++Socratic} triggers clarifying questions to surface hidden assumptions.
\texttt{+++Planning} organizes the clarified elements into a hypothesis schema, while \texttt{+++StepByStep} documents the sequential logic from concept to testable proposition.

\paragraph{Outcome}
This composition guides researchers through explicit reasoning steps, enhancing conceptual rigor and reproducibility in hypothesis formation.

\paragraph{Example}

\begin{tcolorbox}[decoratorverbatim]
\begin{verbatim}
+++Socratic
+++Planning
+++StepByStep

Help refine my idea about how feedback frequency affects team creativity.
\end{verbatim}
\end{tcolorbox}

\textbf{Output (abridged):}

\begin{tcolorbox}[decoratorverbatim]
\begin{verbatim}
[Socratic Inquiry]
What constitutes feedback quality? Who delivers it? How is creativity measured?

[Plan]
1. Define variables: feedback frequency (independent), creativity (dependent).
2. Identify population: design teams in innovation projects.
3. Choose measure: expert-rated creativity scores.

[Step-by-Step Reasoning]
Excessive feedback may reduce autonomy; infrequent feedback limits learning.
Optimal creativity likely occurs under moderate feedback conditions.

[Hypothesis]
Teams receiving biweekly feedback generate higher creativity scores
than those receiving daily or monthly feedback.
\end{verbatim}
\end{tcolorbox}

\subsection{Constructive Feedback Composer}

\textbf{Decorators:} \texttt{+++Candor}, \texttt{+++Tone}, and \texttt{+++Refine(iterations=2)}

\paragraph{Context and Pain Point}
Leaders often struggle to deliver feedback that is honest yet supportive.
Overly blunt communication can damage trust, while vague praise can hinder performance clarity.

\paragraph{Solution}
\texttt{+++Candor} ensures feedback remains truthful and specific.
\texttt{+++Tone} enforces professional empathy, and \texttt{+++Refine(iterations=2)} calibrates phrasing through successive edits to balance precision and respect.

\paragraph{Outcome}
The combination produces feedback messages that preserve integrity while minimizing interpersonal friction.

\paragraph{Example}

\begin{tcolorbox}[decoratorverbatim]
\begin{verbatim}
+++Candor
+++Tone(style=professional)
+++Refine(iterations=2)

Compose feedback for a talented team member who often misses deadlines.
\end{verbatim}
\end{tcolorbox}

\textbf{Output (abridged):}

\begin{tcolorbox}[decoratorverbatim]
\begin{verbatim}
[Candor]
Your technical work is strong, but delays have affected team delivery.

[Refinement 1]
You bring valuable expertise, yet timely delivery will elevate project success.

[Refinement 2]
Your technical contributions significantly strengthen the team.
Improving consistency in deadlines will reinforce your leadership potential.

[Tone: Professional]
Direct yet supportive.
\end{verbatim}
\end{tcolorbox}

\subsection{Research Title Ideation Assistant}

\textbf{Decorators:} \texttt{+++Brainstorm}, \texttt{+++Reasoning}, and \texttt{+++Refine(iterations=3)}

\paragraph{Context and Pain Point}
Selecting an effective research paper title requires balancing precision with engagement.
AI suggestions can be generic or misaligned with disciplinary conventions.

\paragraph{Solution}
\texttt{+++Brainstorm} generates a wide range of creative candidates.
\texttt{+++Reasoning} explains the rationale behind each title’s framing, and \texttt{+++Refine(iterations=3)} filters and polishes the strongest options through guided iteration.

\paragraph{Outcome}
Researchers obtain well-reasoned, field-appropriate titles that improve clarity and visibility without substituting for human authorship.

\paragraph{Example}

\begin{tcolorbox}[decoratorverbatim]
\begin{verbatim}
+++Brainstorm
+++Reasoning
+++Refine(iterations=3)

Generate possible titles for a paper on human–AI collaboration in policy analysis.
\end{verbatim}
\end{tcolorbox}

\textbf{Output (abridged):}

\begin{tcolorbox}[decoratorverbatim]
\begin{verbatim}
[Brainstorm]
1. Shared Sensemaking: Human–AI Synergy in Policy Work
2. Bridging Intuition and Computation: The Analyst’s New Partner
3. Collaborative Intelligence in Decision-Making

[Reasoning]
Title 1 clearly signals scope (sensemaking), method (collaboration),
and domain (policy).

[Refinement 1]
Filter out ambiguous metaphors; prefer explicit academic phrasing.

[Refined Titles]
1. Shared Sensemaking: Human–AI Collaboration in Policy Analysis
2. Collaborative Intelligence for Transparent Policy Decisions
\end{verbatim}
\end{tcolorbox}

\subsection{Prompt Refinement Assistant}

\textbf{Decorators:} \texttt{+++Rewrite}, \texttt{+++Reasoning}

\paragraph{Context and Pain Point}
Users often formulate vague or underspecified prompts, leading to ambiguous or uninformative outputs.
Even small phrasing issues—such as missing context, unclear objectives, or lack of target audience—can degrade model performance.

\paragraph{Solution}
\texttt{+++Rewrite} reformulates the original prompt to improve clarity, specificity, and alignment with the intended goal.
After rewriting, \texttt{+++Reasoning} executes the new prompt and provides explicit reasoning for the final answer, making the response process transparent and interpretable.

\paragraph{Outcome}
The combination improves both prompt quality and model accountability.
It teaches users how refined phrasing alters reasoning pathways, enhancing prompt literacy and consistency of outcomes.

\paragraph{Example}

\begin{tcolorbox}[decoratorverbatim]
\begin{verbatim}
+++Rewrite
+++Reasoning

Explain photosynthesis for a class.
\end{verbatim}
\end{tcolorbox}

\textbf{Output (abridged):}

\begin{tcolorbox}[decoratorverbatim]
\begin{verbatim}
[Rewritten Prompt]
Explain the process of photosynthesis to middle school students,
emphasizing the role of sunlight, chlorophyll, and energy conversion.

[Reasoning]
Photosynthesis converts light energy into chemical energy stored in glucose.
Chlorophyll absorbs sunlight, enabling plants to transform carbon dioxide and water
into glucose and oxygen through a series of reactions in the chloroplasts.

[Final Answer]
Photosynthesis is how plants make their own food using sunlight.
Inside their leaves, chlorophyll captures sunlight, turning carbon dioxide and water
into glucose (a type of sugar) and releasing oxygen into the air.
\end{verbatim}
\end{tcolorbox}

\subsection{Cross-Case Insights and Design Principles}

Across these applications, prompt decorators demonstrate compositional reasoning: each combination couples a \textit{cognitive control} decorator (e.g., \texttt{+++Debate}, \texttt{+++Socratic}) with an \textit{expression or refinement} decorator (e.g., \texttt{+++Refine}, \texttt{+++Tone}). 
This pairing enables human–AI collaboration characterized by transparency, consistency, and ethical alignment.

\begin{table}[h!]
\centering
\caption{Decorator Combinations and Their Reasoning Archetypes}
\begin{tabular}{p{5cm} p{8cm}}
\toprule
\textbf{Decorator Composition} & \textbf{Reasoning Archetype} \\
\midrule
\texttt{+++Debate} \newline \texttt{+++Reasoning} \newline \texttt{+++Refine} 
& Dialectical deliberation and convergence toward a balanced decision \\
\midrule
\texttt{+++Socratic} \newline \texttt{+++Planning} \newline \texttt{+++StepByStep} 
& Inquiry-driven logical structuring of research hypotheses \\
\midrule
\texttt{+++Candor} \newline \texttt{+++Tone} \newline \texttt{+++Refine} 
& Emotionally intelligent communication and tone calibration \\
\midrule
\texttt{+++Brainstorm} \newline \texttt{+++Reasoning} \newline \texttt{+++Refine} 
& Divergent–convergent ideation through analytical filtering \\
\midrule
\texttt{+++Rewrite} \newline \texttt{+++Reasoning} 
& Prompt improvement followed by transparent reasoning in executing the refined instruction \\
\bottomrule
\end{tabular}
\end{table}

\paragraph{Closing Reflection}
These cases demonstrate that composable decorators provide more than stylistic control—they instantiate epistemic scaffolding.
By encoding reasoning modes such as debate, inquiry, and refinement, decorators transform language models from content generators into structured reasoning assistants that support transparent cognitive workflows.


\section{Risks and Limitations}
\label{sec:risks}

Prompt Decorators introduce a declarative and composable layer of reasoning, formatting, and behavioral control for large language models (LLMs). 
However, they also entail risks stemming from the probabilistic nature of LLMs and human factors in prompt design. 
Recognizing these constraints is essential for responsible implementation.

\subsection{Interpretive Ambiguity and Behavioral Drift}

Because decorators rely on pattern recognition rather than deterministic parsing, their behavior can vary across sessions or contexts. 
For example, \texttt{+++StepByStep} may produce inconsistent granularity, while \texttt{+++Reasoning} may blend logic with conclusion. 
This interpretive drift limits semantic precision and reproducibility. Declarative syntax alone cannot ensure predictability. Middleware or symbolic parsing may be required for deterministic outcomes.

\subsection{Overreliance on Simulated Reasoning}

Decorators like \texttt{+++Reasoning}, \texttt{+++Debate}, and \texttt{+++Critique} often generate performative rather than genuine reasoning. 
They simulate internal logic without revealing real inference paths, creating an illusion of interpretability. Expressed reasoning improves clarity but not epistemic reliability; users must avoid mistaking rhetorical structure for cognitive transparency.

\subsection{Decorator Conflicts and Cascading Effects}

Composing decorators such as \texttt{+++Refine}, \texttt{+++Tone}, and \texttt{+++OutputFormat} can cause interference—formatting may truncate reasoning, or refinement may override tone. 
Without precedence rules, such compositions can cascade unpredictably. Complex combinations require controlled execution pipelines or dependency management to prevent behavioral conflicts.

\subsection{Usability and Cognitive Overhead}

While decorators simplify control, their syntax introduces conceptual load. 
Non-technical users may find constructs like \texttt{+++Decorator(parameter=value)} unintuitive, diverting attention from content creation. User-friendly interfaces, autocomplete tools, or GUI-based selectors are necessary for adoption beyond technical audiences.

\subsection{Ethical and Governance Risks}

Decorators affecting tone or candor, like \texttt{+++Candor(level=high)} or \texttt{+++Tone(style=persuasive)}, may unintentionally shape bias or manipulative phrasing. 
Persistent decorators under \texttt{+++ChatScope} can propagate misconfigurations across sessions. Governance mechanisms should log decorator metadata and enforce transparency, neutrality, and auditability.

\subsection{Model and Architecture Dependency}

Different models or fine-tuning contexts interpret declarative syntax inconsistently. 
Open-weight systems may treat decorators as plain text, while tuned models may overfit to patterns. Cross-model portability demands standardized interpreter layers or adaptation modules.

\subsection{Hallucination and Consistency Risks}

Reasoning-oriented or iterative decorators (\texttt{+++Reasoning}, \texttt{+++Refine}) may amplify hallucinations if early errors propagate through refinement cycles. Refinement should be paired with factual verification or retrieval mechanisms, especially in high-stakes tasks.

\subsection{Security and Injection Vulnerabilities}

Decorator syntax can be exploited via prompt injection—malicious inputs may insert fake \texttt{+++Clear} or \texttt{+++ChatScope} commands to disable safeguards. Secure implementations must sandbox parsing, sanitize inputs, and isolate user from system-level decorators.

\subsection{Evaluation and Benchmarking Gaps}

Since behavior depends on semantic interpretation, benchmarking decorator efficacy requires human judgment, introducing subjectivity. Standardized metrics for reasoning coherence and stylistic control are needed to assess effectiveness objectively.

\subsection{Versioning and Maintenance Challenges}

As new decorators emerge, overlap and incompatibility may occur across sessions or model versions. 
Persistent decorators can behave differently after updates, reducing reproducibility. Long-term stability requires version-controlled registries and consistent documentation across model generations.


\section{Future Work and Possible Extensions}
\label{sec:future-work}

The Prompt Decorator framework establishes a foundation for declarative and composable control in Large Language Models (LLMs). While it advances transparency, structure, and modularity, several research directions remain open—ranging from standardization to adaptive, context-aware control.

\subsection{Standardization and Execution Middleware}
A unified Prompt Decorator Specification (PDS) should define naming conventions, parameters, and precedence rules to ensure consistent behavior across models and platforms. Achieving such interoperability will likely require middleware capable of interpreting decorators before prompt delivery. This layer could resolve dependencies, enforce precedence, and validate outputs—turning symbolic directives into executable behavioral constraints. Shared registries and standardized interpreter APIs would make decorators portable across model architectures while maintaining deterministic control.

\subsection{Integration with Agent Frameworks}
Prompt Decorators could function as a behavioral grammar for multi-agent systems, defining reasoning styles and coordination strategies. Agents may interpret decorators as dynamic behavioral contracts—such as “plan,” “critique,” or “debate”—to enable self-configuring and auditable reasoning across distributed AI environments.

\subsection{Evaluation and Benchmarking}
Future research should develop standardized benchmarks that assess how decorators affect reasoning quality, coherence, creativity, and alignment. Metrics such as behavioral fidelity, consistency under ChatScope, and compositional robustness would enable systematic evaluation and fine-tuning.

\subsection{Hybrid Declarative–Procedural Interfaces}
Combining decorators with procedural frameworks like LMQL, BAML, or DSPy could bridge natural-language control and programmatic logic. Decorators might trigger structured code blocks or generate execution graphs that compile back into readable syntax, blending flexibility with formal rigor.

\subsection{Adaptive and Context-Aware Decorators}
Next-generation decorators may adjust dynamically to context or user intent. For example, \texttt{+++Tone} could shift from “technical” to “empathetic” based on sentiment, while \texttt{+++Reasoning} could adapt verbosity automatically. Feedback-driven tuning and meta-decorators would enable self-optimizing behavioral primitives.

\subsection{Ethical and Governance Considerations}
Since decorators can influence tone and reasoning style, governance mechanisms must ensure transparency and neutrality. Audit logs, bias analyses, and metadata tracking should document which decorators influenced an output and how. Ethical safeguards must guarantee that declarative control enhances accountability rather than manipulation.

\subsection{Long-Term Vision}
Over time, decorators could evolve into a shared declarative language for reasoning and control, comparable to HTML for structure or SQL for data. Such a standard would enable portable, self-documenting cognitive workflows where identical decorator chains behave consistently across models, creating a transparent interface between human intent and machine reasoning.


\section{Conclusion}
\label{sec:conclusion}

Prompt Decorators provide a structured, declarative framework for shaping the reasoning, tone, and structure of large language model (LLM) outputs. 
By separating intent from content and introducing composable behavioral controls, they enable users to specify not only \textit{what} to produce but also \textit{how} to produce it. 
This separation creates a modular interface between human intent and model behavior, offering a reproducible method for achieving consistent and interpretable results.

The framework’s strength lies in its flexibility: decorators can be composed to build tailored reasoning chains, enforce stylistic norms, or introduce multi-step refinement processes. 
Their declarative nature also supports auditability and transparency, two features increasingly vital for responsible AI deployment. 
When implemented carefully, Prompt Decorators help shift prompting from an ad-hoc art toward a more structured engineering practice.

However, the concept remains in an exploratory stage. 
Interpretive ambiguity, behavioral drift, and interaction conflicts still pose practical challenges. 
Decorator performance depends heavily on model architecture, fine-tuning data, and prompt parsing fidelity. 
Additionally, without proper governance, tone and reasoning decorators may amplify bias or produce manipulative framing.

Future research should focus on standardization, interpreter-based enforcement, and evaluation benchmarks to ensure consistency and ethical reliability. 
Integrating decorators with agent frameworks, procedural logic, or adaptive systems could extend their capabilities further—potentially establishing a universal declarative layer for human–AI reasoning.

Ultimately, Prompt Decorators represent a step toward a more transparent and governable paradigm of language model interaction: one where structure, reasoning, and style are no longer implicit model behaviors, but explicit, composable components under human control.


\newpage

\bibliographystyle{IEEEtran}

\begin{thebibliography}{10}
\providecommand{\natexlab}[1]{#1}
\providecommand{\url}[1]{\texttt{#1}}
\expandafter\ifx\csname urlstyle\endcsname\relax
  \providecommand{\doi}[1]{doi: #1}\else
  \providecommand{\doi}{doi: \begingroup \urlstyle{rm}\Url}\fi

\bibitem[Beurer-Kellner et~al.(2023)Beurer-Kellner, Fischer, and Vechev]{beurerkellner2023lmql}
Luca Beurer-Kellner, Marc Fischer, and Martin Vechev.
\newblock Prompting is programming: A query language for large language models.
\newblock In \emph{Proceedings of the ACM on Programming Languages}, 2023.
\newblock \doi{10.1145/3591300}.

\bibitem[Bommasani et~al.(2021)Bommasani, Hudson, Adeli, et~al.]{bommasani2021opportunities}
Rishi Bommasani, Drew~A. Hudson, Ehsan Adeli, et~al.
\newblock On the opportunities and risks of foundation models.
\newblock \emph{arXiv preprint arXiv:2108.07258}, 2021.

\bibitem[{BoundaryML}(2024)]{boundaryml2024baml}
{BoundaryML}.
\newblock Baml: Behavioral api modeling language.
\newblock \url{https://docs.boundaryml.com/}, 2024.
\newblock Accessed: October 2025.

\bibitem[Khattab et~al.(2022)Khattab, Santhanam, Li, Hall, Liang, Potts, and Zaharia]{khattab2022demonstrate}
Omar Khattab, Keshav Santhanam, Xiang~Lisa Li, David Hall, Percy Liang, Christopher Potts, and Matei Zaharia.
\newblock Demonstrate-search-predict: Composing retrieval and language models for knowledge-intensive {NLP}.
\newblock \emph{arXiv preprint arXiv:2212.14024}, 2022.

\bibitem[Khattab et~al.(2024)Khattab, Singhvi, Maheshwari, Zhang, Santhanam, Vardhamanan, Haq, Sharma, Joshi, Moazam, Miller, Zaharia, and Potts]{khattab2024dspy}
Omar Khattab, Arnav Singhvi, Paridhi Maheshwari, Zhiyuan Zhang, Keshav Santhanam, Sri Vardhamanan, Saiful Haq, Ashutosh Sharma, Thomas~T. Joshi, Hanna Moazam, Heather Miller, Matei Zaharia, and Christopher Potts.
\newblock Dspy: Compiling declarative language model calls into self-improving pipelines.
\newblock 2024.

\bibitem[Schulhoff et~al.(2025)Schulhoff, Ilie, Balepur, et~al.]{schulhoff2025promptreport}
Sander Schulhoff, Michael Ilie, Nishant Balepur, et~al.
\newblock The prompt report: A systematic survey of prompt engineering techniques.
\newblock \emph{arXiv preprint arXiv:2406.06608}, 2025.

\bibitem[Vatsal and Dubey(2024)]{vatsal2024survey}
Shubham Vatsal and Harsh Dubey.
\newblock A survey of prompt engineering methods in large language models for different nlp tasks.
\newblock \emph{arXiv preprint arXiv:2407.12994}, 2024.

\bibitem[Wang et~al.(2022)Wang, Wei, et~al.]{wang2022selfconsistency}
Xuezhi Wang, Jason Wei, et~al.
\newblock Self-consistency improves chain-of-thought reasoning in language models.
\newblock \emph{arXiv preprint arXiv:2203.11171}, 2022.

\bibitem[Wei et~al.(2022)Wei, Wang, et~al.]{wei2022chainofthought}
Jason Wei, Xuezhi Wang, et~al.
\newblock Chain-of-thought prompting elicits reasoning in large language models.
\newblock \emph{Advances in Neural Information Processing Systems}, 35:\penalty0 24824--24837, 2022.

\bibitem[Yao et~al.(2023)Yao, Zhao, et~al.]{yao2022react}
Shunyu Yao, Jeffrey Zhao, et~al.
\newblock React: Synergizing reasoning and acting in language models.
\newblock In \emph{International Conference on Learning Representations (ICLR)}, 2023.

\end{thebibliography}

\end{document}